\documentclass[a4paper, 11pt]{article}

\usepackage{graphicx}
\usepackage{lineno}
\usepackage{amssymb}
\usepackage{amssymb,latexsym,amsmath, graphicx, amsfonts, caption, color, verbatim, booktabs}
\usepackage{bbm}
\usepackage[bbgreekl]{mathbbol}

\RequirePackage[OT1]{fontenc}
\RequirePackage[numbers, compress]{natbib}
\usepackage{subcaption}
\numberwithin{equation}{section}

\title{Survival trees for right-censored data based on score based parameter instability test}

\author{Madan Gopal Kundu\thanks{Corresponding author: Madan G Kundu, madan.kundu@abbvie.com.
This article reflects the views of the author and should not be construed to represent AbbVie's views or policies.}\\
AbbVie Inc, North Chicago, IL, USA\\
\\
Samiran Ghosh\\
Department of Family Medicine and Public Health Sciences,\\
Wayne State University School of Medicine, Detroit, MI 48202}
\date{\today}

\begin{document}
\maketitle

\begin{abstract}
Survival analysis of right censored data arises often in many areas of research including medical research. Effect of covariates (and their interactions) on survival distribution can be studied through existing methods which requires to pre-specify the functional form of  the covariates including their interactions. Survival trees offer relatively flexible approach when the form of covariates' effects is unknown. Most of the currently available survival tree construction techniques are not based on a formal test of significance; however, recently proposed ctree algorithm (\cite{hothorn2006unbiased}) uses permutation test for splitting decision that may be conservative at times. We consider parameter instability test of statistical significance of  heterogeneity  to guard against spurious findings of variation in covariates’ effect without being overly conservative. We have proposed SurvCART algorithm to construct survival tree under conditional inference framework (\cite{hothorn2006unbiased}) that selects splitting variable via parameter instability test and subsequently finds the optimal split based on some maximally chosen statistic. 
Notably, unlike the existing algorithms which focuses only on heterogeneity in event time distribution, the proposed SurvCART algorithm can take splitting decision based in censoring distribution as well along with  heterogeneity in event time distribution.
The operating characteristics of parameter instability test and comparative assessment of SurvCART algorithm were carried out via simulation. 
Finally, SurvCART algorithm was applied to a real data setting. The proposed method is fully implemented in  R package LongCART available on CRAN.
\end{abstract}

\section{Introduction}\label{sec:intro}
Application of survival times (or, time-to-event data in general) are numerous and arise in all areas of research. In practice, survival times may be influenced by several covariates. For example, clinical investigators exert a great deal of time and energy in attempts to identify and quantify the effects of prognostic factors, namely, patient characteristics that relate to the course of disease (\cite{davis1989exponential}).  There are parametric and semi-parametric methods (e.g., Cox proportional hazards model) which allows to associate survival time with covariates. However, such models requires to pre-specify the functional form of  the covariates including their interactions. Survival trees offer relatively flexible approach when the form of covariates' effects is unknown and also have a greater ability to detect interactions automatically based on observed data. Survival trees are the non-parametric alternative of (semi-) parametric models and also have the advantage of easier interpretation. Most of the currently available survival tree construction techniques are not based on a formal test of significance and hence may be prone to spurious findings of variation in covariates' effect. In this article we have proposed  a recursive partitioning algorithm to construct survival tree  that selects splitting variable via formal statistical test. Unlike the existing algorithms (\cite{hothorn2006unbiased,therneau1990martingale,  leblanc1992relative, ciampi1988recpam}) which focuses only on heterogeneity in event time distribution, the proposed algorithm provides a framework to identify the subgroups based on heterogeneity in event time and/or censoring time distributions.\\

Tree based method, first introduced by  \citet{morgan1963problems}, is useful in identifying `homogeneous' subgroups defined by some covariates in diverse population. Among the tree based methods, classification and regression tree (CART) methods (\citealp{breiman1984classification}) are the most popular. Recently the concept of CART methodology have been extended in the context of fitting cross-sectional regression models (see, e.g., \citealp{zeileis2008model}) and longitudinal setting (see, e.g., \citealp{ segal1992tree, loh2002regression, sela2012re, loh2013regression, kundu2019regression}). Application of tree approach to survival data can be traced back to \citet{ciampi1981approach} where attempt was made to identify prognostic factors  influencing survival of non-Hodgkin's lymphoma patients.  However, \citet{gordon1985tree} were first to suggest CART paradigm for survival analysis. Since then a number approaches have been proposed for tree construction with survival data (\citealp{davis1989exponential,  leblanc1992relative, ciampi1988recpam,  ciampi1986stratification,  ciampi1987recursive,  segal1988regression,  leblanc1993survival,   ciampi1995tree}); For a structured review of these methods, please refer to  \citet{bou2011review}. \\

The goal of a survival tree (i.e., tree with survival data) is to identify `homogeneous' subgroups, characterized by prognosis variables (i.e. baseline covariates), in a heterogeneous population on the basis of how long they survive, thus enabling classification by prognosis (\citealp{davis1989exponential}). The homogeneity in the context survival  tree refers to ``absence of sufficient statistical evidence of variation in time-to-event distribution". Most of the currently available survival tree construction techniques  are  based on maximally selected statistic  such as  log-rank test statistic (\citealp{ciampi1986stratification, leblanc1993survival}),  Wilcoxon-Gehan statistic (\citealp{ciampi1988recpam}), likelihood ratio statistic (\citealp{ ciampi1987recursive, ciampi1995tree}),  likelihood based deviance (\citealp{leblanc1992relative}), exponential log-likelihood loss (\citealp{davis1989exponential}) or Taron-Ware class of statistic (\citealp{segal1988regression}) for selection of best split. Another approach to construct survival tree is based on martingale residual where martingale residuals from a null Cox model are used as the outcome to construct the tree (\citealp{therneau1990martingale}). However, none of the above mentioned survival tree construction techniques is based on formal statistical test. Lack of formal test of statistical significance in the construction of survival tree may overfit the data (\citealp{leblanc1993survival}) and thus may result in spurious findings (\citealp{negassa2005tree}). To alleviate the issue of spurious findings, some authors have also discussed pruning and amalgamation of trees as well (\citealp{ciampi1995tree}). To avoid the problems associated with the exhaustive search strategies, relatively more recently, \citet{hothorn2006unbiased} proposed a unified conditional inference framework for construction of trees including survival tree. which identifies best split at any given tree node in two steps. In step 1,  a global null hypothesis of independence with response  is tested for each covariate using permutation based test; the  covariate with minimum $p-$ value is chosen as splitting variable, if  found statistically significant. In second step, the split point is chosen based on maximum value of some test statistic. Very recently survival tree algorithms of \citet{hothorn2006unbiased} and \citet{leblanc1993survival} has been extended to left truncated right censored data by \citet{fu2017survival}.\\

We have proposed survival tree construction algorithm to identify subgroups with heterogeneous event time (and, censor time distribution). In our set-up, the homogeneity refers to absence of sufficient statistical evidence of variation in survival curve (and, censoring time distribution). The proposed algorithm is an adaption of LongCART algorithm for constructing tree with longitudinal data  (\citealp{kundu2019regression}) in survival set up.  For the ease of discussion, we name our algorithm as SurvCART algorithm. The proposed SurvCART algorithm works in two steps to identify the best split. In step 1, we first identify whether any partitioning variable  influence survival (and, censoring distribution) via statistical testing. Such tests are often based on score process and commonly known as {\it ``parameter instability test"}. There are several score-based parameter instability test suggested in literature  (\citealp{hothorn2006unbiased, zeileis2008model, kundu2019regression, andrews1993tests, hjort2002tests}). Particularly, \citet{hothorn2006unbiased} considered permutation based test to tests for heterogeneity in construction of survival tree. Although permuation based test does not need specification of underlying distribution,  use of permuation based test has been very limited and controversial particularly due to its conservatism (\citealp{berger2000pros}). In SurvCART algorithm, we  use likelihood based parametric  test to assess homogeneity in parameters of time-to-event distribution (and, censoring distribution).  Parameter instability tests are carried out for each partitioning variable separately and most significant partitioning variable is chosen as splitting variable. If none of the partitioning variables turns out significant in parameter instability testing, the process stops there without further splitting.  Once the splitting variable is chosen, in step 2, the cut-off point with the maximum  test statistic (e.g., Log-rank) for comparing two groups is chosen for binary splitting. The key idea here is that we are combining the multiple testing procedures (step~1) with model selection (step~2) in order to control the type I error while taking the decision on splitting at each node. Such a step minimizes the selection bias in choosing the partitioning variable compared to the exhaustive search-based procedures where the partitioning variables with many unique values tend to have an advantage over the partitioning variables with fewer unique values (\cite{loh2002regression,   loh2013regression}). The SurvCART algorithm provides an improvement over the existing survival tree construction approaches in one or more of the following aspects: (1) the test for the decision about further splitting at each node is type I error controlled via formal parametric hypothesis testing and hence offers guard against variable selection bias, over-fitting and spurious splitting, (2) subgroups are chosen considering heterogeneity both in event time and censoring distribution,  and (4) computational time is greatly reduced.\\

The remainder of this paper is organized as follows. In Section \ref{PrelimLongTree} the survival model of interest are  summarized. Tests for parameter instability  for a single partitioning variable cases are discussed in Section \ref{InstabTest} and its extension to multiple partitioning variables are discussed in Section~\ref{InstabTestMult}. Algorithm for constructing survival trees is discussed in Section \ref{Algorithm}. Results from the simulation studies examining the performance of the instability test and the performance of SurvCART algorithm are reported in Section \ref{SimulationRegTree}. An application example is presented in Section \ref{TreeApplication}. Finally, in Section~\ref{sec:disc}, we discuss the implications of our findings. The SurvCART algorithm is implemented in \verb|SurvCART()| through \verb|R| package  \verb|LongCART| (\citealp{kundu2020longcart}).

\section{Methodology}
\subsection{Notation and preliminaries}\label{PrelimLongTree}
We begin by describing the basic setup which leads to the development of survival
trees. We denote by $T^*$ the true survival time and by $C$ the true censoring
time. The observed data is then composed of $T = \min{(T^*,C)}$, the time until
either the event occurs or the subject is censored; $\delta = I(T^* \le C)$, an indicator
that takes a value of 1 if the true time-to-event is observed and 0 if the subject
is censored. We also assume $T^*$ and $C$ are independent. In addition, for each individual, we observe a vector of $S$ covariates $X=(X_{1}, \ldots, X_{S})$ measured at baseline. We assume that $X_1, \ldots, X_S$ includes all potential baseline attributes that can influence  the either $T^*$ or $C$ or both. Data is available for N independent subjects $(T_i, \delta_i,X_i), i = 1, . . . ,N$. The basic setup assumes that the covariate values are available at time 0 for each subject. Thus, only the baseline values of a time–varying covariate are typically used. It is assumed that the underlying model, generating the data, is consists of $K$ distinct and mutually exclusive terminal subgroups and these subgroups can be characterized by baseline covariates $X_{1}, \ldots, X_{S}$. Further, we assume that, in the $k^{th}$ subgroup, $T^*\sim f(t;\mathbf{\boldsymbol{\theta}}_{Tk})$ and $C\sim g(c; \mathbf{\boldsymbol{\theta}}_{Ck})$. Also, $\mathbf{\boldsymbol{\theta}}_{k}=[\mathbf{\boldsymbol{\theta}}_{Tk}, \mathbf{\boldsymbol{\theta}}_{Ck}]^{\top}$, and $S(\cdot)$ and $H(\cdot)$ are complementary cumulative distribution functions associated with $T^*$ and $C$, respectively. When homogeneity holds for entire population, we have $\mathbf{\boldsymbol{\theta}}_{Tk}\equiv\mathbf{\boldsymbol{\theta}}_{T}$ and  $\mathbf{\boldsymbol{\theta}}_{Ck}\equiv\mathbf{\boldsymbol{\theta}}_{C}$ implying $\mathbf{\boldsymbol{\theta}}_{k}\equiv\mathbf{\boldsymbol{\theta}}$ where $\mathbf{\boldsymbol{\theta}}=[\mathbf{\boldsymbol{\theta}}_T, \mathbf{\boldsymbol{\theta}}_C]^{\top}$. Then the contribution of $i^{th}$ subject in the log-likelihood is
\[
l_i(\boldsymbol{\theta})=[f(t_i;\mathbf{\boldsymbol{\theta}})  H(t_i;\mathbf{\boldsymbol{\theta}})]^{\delta_i}
[S(t_i;\mathbf{\boldsymbol{\theta}})  g(t_i;\mathbf{\boldsymbol{\theta}})]^{1-\delta_i}
\]
The score function for estimating $\boldsymbol{\theta}$ pertaining to $i^{th}$ subject $\frac{\partial }{\partial \boldsymbol{\theta}}l_i(\boldsymbol{\theta})=\mathbf{u}_i(\boldsymbol{\theta})=[\mathbf{u}_i(\boldsymbol{\theta}_T),  \mathbf{u}_i(\boldsymbol{\theta}_C)]^{\top}$ with variance $\mathbf{J(\boldsymbol{\theta})}=\mbox{Var}[\mathbf{u}_i(\boldsymbol{\theta})]=-E[\frac{\partial^2 }{\partial^2 \boldsymbol{\theta}}l_i(\boldsymbol{\theta})]$. Since $T^*$ and $C$ are independent, we have,
\begin{itemize}
\item{C1:} $\mathbf{u}_i(\boldsymbol{\theta}_T)$ does not involve $\boldsymbol{\theta}_C$
\item{C2:} $\mathbf{u}_i(\boldsymbol{\theta}_C)$ does not involve $\boldsymbol{\theta}_T$
\item{C3:} $\mathbf{J(\boldsymbol{\theta})}=\mbox{diag}\{\mathbf{J(\boldsymbol{\theta}_T)}, \mathbf{J(\boldsymbol{\theta}_C)}\}$ since $cov[\mathbf{u}_i(\boldsymbol{\theta}_T), \mathbf{u}_i(\boldsymbol{\theta}_C)]=0$.
\end{itemize}

Further, the maximum likelihood (ML) estimate of $\mathbf{\boldsymbol{\theta}}$ using all the observation from $N$ subjects is $\hat{\mathbf{\boldsymbol{\theta}}}$. The total number of observed events is $D$ and  $S_T=\sum_{i=1}^N{t_i}$ is sum of all the follow-up time. Now, the likelihood estimate $\hat{\mathbf{\boldsymbol{\theta}}}$ is valid only if the entire population under consideration is homogeneous (i.e. $\theta_{k}=\theta$, $\forall k$).  With respect to a given partitioning variable, homogeneity refers to that the true value of $\mathbf{\boldsymbol{\theta}}$ remains the same across all the values of that partitioning variable.
%
\subsection{Test for parameter instability for a single partitioning variable} \label{InstabTest}
The purpose of  {\em parameter instability test} is to test whether the true value of $\mathbf{\boldsymbol{\theta}}$ remains the same across all distinct values of baseline partitioning variable. Let $X\in \{X_1, \ldots, X_S\}$ be any partitioning variable with $G$ ordered cut-off points: $c_{(1)}< \ldots < c_{(G)}$ and $\boldsymbol{\theta}_{(g)}$ be the true value of $\boldsymbol{\theta}$ when $X=c_{(g)}$. Assume that there are $m_g$ subject with  $X=c_{(g)}$. We denote the cumulative number of subjects with $X\leq c_{(g)}$ by $M_g$. That is, $M_g=\sum_{j=1}^g{m_j}$ and  $M_G=\sum_{j=1}^G{m_j}=N$. We want to conduct an omnibus test,
\begin{equation*}
H_0: \boldsymbol{\theta}_{(g)}=\boldsymbol{\theta}_0  \; {\mbox vs.} \;
H_1: \boldsymbol{\theta}_{(g)} \neq \boldsymbol{\theta}_0.
\end{equation*}
Here, $H_0$ indicates the scenario when parameter $\boldsymbol{\theta}$ remains constant (that is, {\it homogeneity}) at some common value $\boldsymbol{\theta}_0=[\boldsymbol{\theta}_{T0}, \boldsymbol{\theta}_{C0}]^{\top}$ and $H_1$ corresponds to the situation of parameter instability (that is, {\it heterogeneity}). 
In the two subsections to follow, we have summarized the parameter instability test depending on whether the partitioning variable $X$ is categorical or continuous. These tests are formulated following \citet{kundu2019regression} and details are  given in Appendix.  
%
%
\subsubsection{Instability test with categorical partitioning variable}\label{sec:StabCat}
When the  partitioning  variable $X$ is  categorical with a small number of categories (that is, $G \ll N$), following test statistics
\begin{align}
\chi^2_T&=\sum_{g=1}^G{\left[\sum_{i=1}^N{I(X_i=c_{(g)})\mathbf{u}(\mathbf{y}_i, \hat{\boldsymbol{\theta}}_T)}\right]^{\top}\left[m_g\mathbf{J}(\hat{\boldsymbol{\theta}}_T)\right]^{-1}\left[\sum_{i=1}^N{I(X_i=c_{(g)})\mathbf{u}(\mathbf{y}_i, \hat{\boldsymbol{\theta}}_T)}\right]} \label{eq:stabcatT}\\
\chi^2_C&=\sum_{g=1}^G{\left[\sum_{i=1}^N{I(X_i=c_{(g)})\mathbf{u}(\mathbf{y}_i, \hat{\boldsymbol{\theta}}_C)}\right]^{\top}\left[m_g\mathbf{J}(\hat{\boldsymbol{\theta}}_C)\right]^{-1}\left[\sum_{i=1}^N{I(X_i=c_{(g)})\mathbf{u}(\mathbf{y}_i, \hat{\boldsymbol{\theta}}_C)}\right]} \label{eq:stabcatC}
\end{align}
are asymptotically distributed as $\chi^2$ with $\dim(\boldsymbol{\theta}_T)\cdot(G-1)$ and $\dim(\boldsymbol{\theta}_C)\cdot(G-1)$ degrees of freedom, respectively.  Here, $I(\cdot)$ is the indicator function. Details are provided in Appendix A. 

%
%
\subsubsection{Instability test with continuous partitioning variable}\label{sec:StabCont}
When $X$  is continuous, consider the following standardized estimated score process
\begin{align}
 \mathbf{M}_N(t, \boldsymbol{\theta}_T)&=N^{-1/2}\mathbf{J}^{-1/2}(\hat{\boldsymbol{\theta}}_T) \sum_{i=1}^{M_g}{\mathbf{u}(\mathbf{y}_i, \boldsymbol{\theta}_T)} \qquad t\in [t_g, t_{g+1}),\\
  \mathbf{M}_N(t, \boldsymbol{\theta}_C)&=N^{-1/2}\mathbf{J}^{-1/2}(\hat{\boldsymbol{\theta}}_C) \sum_{i=1}^{M_g}{\mathbf{u}(\mathbf{y}_i, \boldsymbol{\theta}_C)} \qquad t\in [t_g, t_{g+1})
\end{align}
with $t_g=\dfrac{M_g}{N}$. Then, as shown in Appendix B, the following test statistics corresponding to individual components of $\boldsymbol{\theta}_T=\{\theta_{T,q}; q=1,\cdots, \dim{(\boldsymbol{\theta}_T)}\}$ and $\boldsymbol{\theta}_C=\{\theta_{C,r}; r=1,\cdots, \dim{(\boldsymbol{\theta}_C)}\}$ 
\begin{equation}\label{eq:D_T}
D(\theta_{T,q})\equiv \max_{0 \le t \le 1}{|M_N(t, \hat{\theta}_{T,q})|} = \max_{1 \le j \le N-1}{ |M_N(t, \hat{\theta}_{T,q})|},
\end{equation}
\begin{equation}\label{eq:D_C}
D(\theta_{C,r})\equiv \max_{0 \le t \le 1}{|M_N(t, \hat{\theta}_{C,r})|} = \max_{1 \le j \le N-1}{ |M_N(t, \hat{\theta}_{C,r})|}
\end{equation}
are independent and, under $H_0$, are asymptotically distributed with distribution function (\citealp{billingsley2009convergence})
\begin{equation}\label{eq:D_dist}
F_D(x)=1+2\sum_{l=1}^{\infty}{(-1)^l\exp{(-2\;l^2x^2)}}.
\end{equation}
where, $F_D(\cdot)$ represents cumulative distribution function of the supremum of standard Brownian Bridge process. Clearly, If $\dim(\boldsymbol{\theta})>1$, then the parameter instability test for a single partitioning variable involves multiple testing simultaneously, one test for each parameter of $\boldsymbol{\theta})$. In this case, these p-values need to be adjusted using Hochberg's procedure (\citealp{hochberg1988sharper}) or other similar procedure  and minimum of these two adjusted p-values should be regarded as the overall p-value corresponding to the partitioning variable $X$ (\citealp{dmitrienko2009multiple}).

\subsubsection{Special case under some know parametric distributions} \label{sec:expwei}

The parameter instability test presented in Section~\ref{sec:StabCat} and \ref{sec:StabCont} requires MLEs ($\hat{\boldsymbol{\theta}}_T$ and $\hat{\boldsymbol{\theta}}_C$),   score functions ($\mathbf{u}_i(\boldsymbol{\theta}_T)$ and $\mathbf{u}_i(\boldsymbol{\theta}_C)$) and the variance of score functions ($\mbox{Var}[\mathbf{u}_i(\boldsymbol{\theta}_T)]$ and $\mbox{Var}(\mathbf{u}_i[\boldsymbol{\theta}_C)]$. Here, we specifically present the the parameter instability test under exponential and weibull distributions and have outlined the approach for other including complex distributions. Under exponential distribution, test statistics for parameter instability tests discussed in Section~\ref{sec:StabCat} and \ref{sec:StabCont}  are simplified in great extent as shown below. Please note that even though, for ease of discussion, we have considered below $T^*$ and $C$ follows similar distributions,  it is not mandatory to have similar distribution for both $T^*$ and $C$. For example, it is perfect to have $T^*$ to follow weibull while $C$ to follow log-normal distribution. \\

\underline{Exponential distribution: $f(t;\mathbf{\boldsymbol{\theta}}_{T})=\mbox{Exponential}(\lambda_T)$ and $g(c; \mathbf{\boldsymbol{\theta}}_{C})=\mbox{Exponential}(\lambda_C)$}

\[
f(t;\mathbf{\boldsymbol{\theta}}_{T})=\lambda_T\exp{(-\lambda_T t)}
\qquad
g(c;\mathbf{\boldsymbol{\theta}}_{C})=\lambda_C\exp{(-\lambda_C t)}
\]
The score functions are as follows: 
\begin{equation*}
\mathbf{u}_i(\lambda_T)=\frac{\delta_i}{\lambda_T} - t_i
\qquad
\mathbf{u}_i(\lambda_C)=\frac{1-\delta_i}{\lambda_C} - t_i
\end{equation*}

Based on these score functions, $\hat{\lambda}_T=D/S_T$ and $\hat{\lambda}_C=(N-D)/S_T$ are ML estimators. Further,

\begin{equation*}
\mathbf{J}(\lambda_T)=\frac{D}{N}\cdot\lambda_T^{-2}
\qquad
\mathbf{J}(\lambda_C)=\frac{N-D}{N}\cdot\lambda_C^{-2}
\end{equation*}

With this, for parameter instability test corresponding to categorical partitioning variable, the test statistics in Eq. \eqref{eq:stabcatT} and \eqref{eq:stabcatC} can be simplified as follows:
\[
\chi^2_{T}=\frac{N}{D}\sum_{g=1}^G{\frac{1}{m_g} (d_g - \hat{\lambda}_T\cdot s_g)^2  }
\qquad
 \chi^2_{C}=\frac{N}{N-D}\sum_{g=1}^G{\frac{1}{m_g}  
  (m_g - d_g - \hat{\lambda}_C\cdot s_g)^2}
\]

where, $d_g=\sum_{i=1}^N{I(X_i=c_{(g)})  \delta_i}$ is the number of events and $s_g=\sum_{i=1}^N{I(X_i=c_{(g)})  T_i}$ is the sum of observed follow-up times among the subjects with $X=c_{(g)}$. Both of the above test statistics are asymptotically distributed as $\chi^2$ with $G-1$ degrees of freedom.\\

Further, for parameter instability test corresponding to categorical partitioning variable, the test statistics in Eq. \eqref{eq:D_T} and \eqref{eq:D_C} can be simplified as follows:
\begin{equation} \label{eq:D_T_exp}
D(\lambda_T)=D^{-1/2}\max_{1 \le g \le G-1}{|D_g - \lambda_T\cdot S_g|}
\end{equation}
\[
D(\lambda_C)=(N-D)^{-1/2}\max_{1 \le g \le G-1}{|M_g-D_g - \lambda_C\cdot S_g|}
\]
where, $D_g=\sum_{i=1}^N{I(X_i\le c_{(g)})  \delta_i}$ is the number of events and $S_g=\sum_{i=1}^N{I(X_i\le c_{(g)})  T_i}$ is the sum of observed follow-up times among the subjects with $X \le c_{(g)}$. \\

\underline{Weibull distribution: $f(t;\mathbf{\boldsymbol{\theta}}_{T})=\mbox{Weibull}(\alpha_T, \lambda_T)$ and $g(c; \mathbf{\boldsymbol{\theta}}_{C})=\mbox{Weibull}(\alpha_C, \lambda_C)$}

\[
f(t;\mathbf{\boldsymbol{\theta}}_{T})=\alpha_T\lambda_T t^{\alpha_T-1}\exp{(-\lambda_T t^{\alpha_T})}
\qquad
g(c;\mathbf{\boldsymbol{\theta}}_{C})=\alpha_C\lambda_C t^{\alpha_C-1}\exp{(-\lambda_C t^{\alpha_C})}
\]
The score functions are as follows: 
\begin{equation*}
\mathbf{u}_i(\boldsymbol{\theta}_T)=
\left[
\begin{matrix}
u_i(\alpha_T)\\
u_i(\lambda_T)\\
\end{matrix}
\right]
=
\left[
\begin{matrix}
\frac{\delta_i}{\alpha_T} + \delta_i \log{t_i} - \lambda_T  t_i^{\alpha_T} \log{t_i}\\
\frac{\delta_i}{\lambda_T} - t_i^{\alpha_T}\\
\end{matrix}
\right]
\end{equation*}

\begin{equation*}
\mathbf{u}_i(\boldsymbol{\theta}_C)=
\left[
\begin{matrix}
u_i(\alpha_C)\\
u_i(\lambda_C)\\
\end{matrix}
\right]
=
\left[
\begin{matrix}
\frac{1-\delta_i}{\alpha_C} + (1-\delta_i) \log{t_i} - \lambda_C  t_i^{\alpha_C} \log{t_i}\\
\frac{1-\delta_i}{\lambda_C} - t_i^{\alpha_C}\\
\end{matrix}
\right]
\end{equation*}

with corresponding variances as follows:

\begin{equation*}
\mathbf{J}(\boldsymbol{\theta}_T)=
\left[
\begin{matrix}
\frac{D}{N}\lambda_T^{-2} &
\frac{1}{N}\sum\limits_{i=1}^N{t_i^{\alpha_T} \log{t_i}} \\
\frac{1}{N}\sum\limits_{i=1}^N{t_i^{\alpha_T} \log{t_i}} & 
\frac{D}{N}\alpha_T^{-2} +
\lambda_T  \frac{1}{N}\sum\limits_{i=1}^N{t_i^{\alpha_T}(\log{t_i})^2}\\
\end{matrix}
\right]
\end{equation*}

\begin{equation*}
\mathbf{J}(\boldsymbol{\theta}_C)=
\left[
\begin{matrix}
(1-\frac{D}{N})\lambda_C^{-2} &
\frac{1}{N}\sum\limits_{i=1}^N{t_i^{\alpha_C} \log{t_i}} \\
\frac{1}{N}\sum\limits_{i=1}^N{t_i^{\alpha_C} \log{t_i}}  & 
(1-\frac{D}{N})\alpha_C^{-2} +
\lambda_C  \frac{1}{N}\sum\limits_{i=1}^N{t_i^{\alpha_C}(\log{t_i})^2}\\
\end{matrix}
\right]
\end{equation*}

Note that the ML estimators $\hat{\lambda}_T$, $\hat{\alpha}_T$, $\hat{\lambda}_C$ and $\hat{\alpha}_C$ have to be obtained iteratively. Now,  the test statistics in \eqref{eq:stabcatT}, \eqref{eq:stabcatC}, \eqref{eq:D_T} and \eqref{eq:D_C} can be obtained plugging these expressions. \\

For other survival distributions the expression for MLEs, score functions and variance of score function can be obtained similarly as presented above. Unfortunately, it is not  straightforward to obtain these expressions for some distributions (e.g., Log-normal). However,  for many of  these distributions, MLEs $\hat{\boldsymbol{\theta}}_T$ and $\hat{\boldsymbol{\theta}}_C$  along with $\mbox{Var}(\mathbf{u}_i(\boldsymbol{\theta}_T))$ and $\mbox{Var}(\mathbf{u}_i(\boldsymbol{\theta}_C))$ can be obtained from standard softwares. With this, $\mathbf{J}(\boldsymbol{\theta}_T)$ and $\mathbf{J}(\boldsymbol{\theta}_C)$ can be easily obtained as follows: $\mathbf{J}(\boldsymbol{\theta}_T)=\mbox{Var}[\mathbf{u}_i(\boldsymbol{\theta}_T)]=\frac{1}{N}\mbox{Var}^{-1}[\hat{\boldsymbol{\theta}}_T]$ and $\mathbf{J}(\boldsymbol{\theta}_C)=\mbox{Var}[\mathbf{u}_i(\boldsymbol{\theta}_C)]=\frac{1}{N}\mbox{Var}^{-1}[\hat{\boldsymbol{\theta}}_C]$. One still have to obtain the expressions for score functions; however undubtedly that is a much simpler task  and can be obtained relatively easily.  Expressions of score function under log-normal and normal distribution is presented in Appendix C.

\subsection{Extending for parameter instability for a multiple partitioning variables} \label{InstabTestMult}
In practice, we have multiple candidate partitioning variables. Let there be $S$ partitioning variables: $\{X_1, \ldots, X_S\}$. Here the parameter instability test needs to be repeated for each candidate partitioning variables and the p-values corresponding to the individual partitioning variables should be adjusted using Hochberg's procedure (\citealp{hochberg1988sharper}) or other similar procedure to maintain the overall type-I error at each split. The partitioning variable with minimum adjusted p-value should be selected for the splitting provided it is smaller than the overall type I error $\alpha$. The advantage of $p$-value approach is that it offers unbiased partitioning variable selection when the partitioning variables are measured at different scales (\citealp{ hothorn2006unbiased}).

\subsection{Selecting of cut-off point of splitting variable}\label{Splitpoint}
Once the splitting variable is selected, the split point can be identified based on any maximally chosen statistic such as log-rank,  Wilcoxon-Gehan statistic, likelihood based deviance or exponential log-likelihood loss. However, Log-rank statistic seems to be most popular choice by far. \citet{gordon1985tree} suggested the possibility of using the logrank statistic for splitting and also has been used in other survival tree algorithm (e.g., \citealp{leblanc1993survival}). Use of log-rank test leads to a split which assures the best separation of the median survival times (\citealp{ciampi1986stratification}).  Further, log-rank statistic can be represented as linear function (and hence it is to update log-rank test statistic value at a given split point if it's value is known at previous splitting point); and, the log-rank statistic is  stable (i.e. not highly variable) in presence of censoring (\citealp{leblanc1993survival}). \\

Remainder of this paragraph is described assuming log-rank statistic is used for identification of split point. If the selected splitting variable has stronger evidence of heterogeneity in event time distribution (i.e. $p-$value from testing $\phi_T$ is smaller than $p-$value from testing $\phi_C$) then the standard log-rank test is carried out at each splitting point. However, if the selected splitting variable has stronger evidence of heterogeneity in censoring time (i.e. $p-$value from testing $\phi_C$ is smaller than $p-$value from testing $\phi_T$) distribution then the log-rank test is carried out to compare censoring distribution (i.e. considering censoring as event) at each splitting point. In either case, the best splitting point, $c^*$, is the split such that
\[
LR(c^*)=\max_{c\in S_c}{LR(c)}
\]
where, $LR(c)$ is the standardized two-sample log-rank test statistic at split point $c$ and $S(c)$ is the set of all split point of the splitting variable.


\section{Construction of tree: SurvCART Algorithm}\label{Algorithm}

The proposed SurvCART algorithm constructs survival tree in following steps:
\begin{description}
\item {{\bf Step 1. (Selection of splitting variable)}} Perform the parameter instability test for each candidate partitioning variables as explained in section~\ref{InstabTest}. Stop if no partitioning variable is significant at level $\alpha$. Otherwise, choose the partitioning variable with the smallest $p$-value and proceed to Step 2. 
\item {{\bf Step 2. (Selection of splitting point)}} Consider all cut-off points of the chosen covariate. At each cut-off point, calculate the log-rank statistics value (or other maximally chosen statistic, if log-rank statistic is not appropriate). If the censoring distribution was found more heterogeneous compared to time to event distribution for the chosen partitioning variable selected in previous step, compute the logrank test statistic assuming censoring as event; otherwise compute the regular logrank test statistic. Choose the cut-off value that provides the maximum value (see Section~\ref{Splitpoint}).
\item {{\bf Step 3.}} Follow the Steps 1-2 until no covariate founds to be significant through instability test.
\end{description}

\begin{figure}
\centerline{%
\includegraphics [angle=0,width=160mm, height=80mm]{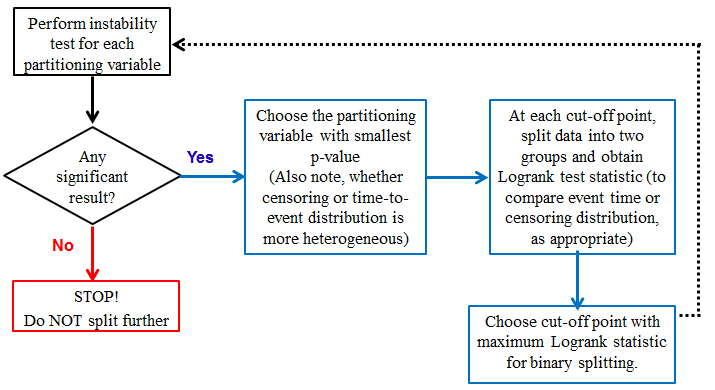}}
\caption{SurvCART algorithm}
\label{fig:algorithm}
\end{figure}
	
The algorithm is also displayed in Figure~\ref{fig:algorithm}. Improvement in survival tree can be quantified using the likelihood based criteria such as AIC and decision on pruning of tree can be driven by adding some penalty for every branch added to the tree as discussed in (\citealp{kundu2019regression}).\\

An important question may arise when constructing survival tree using SurvCART algorithm about the choice of the distributions for $T^*$ and $C$. In absence of any prior knowledge of distribution we recommend either using very flexible distribution such as Weibull distribution. Another recommendation is to construct tree assuming several distributions for $T^*$ and $C$ and then compare the log-likelihood or AIC of the final tree to select the distributions for $T^*$ and $C$. We have compared the AIC of survival tree from SurvCART algorithm with correctly specified model against the mis-specified model (Section~\ref{SimulationRegTree2}). The simulation results presented in Section~\ref{SimulationRegTree2} also suggests that SurvCART algorithm performs relatively better than the other method, even with mis-specified time to event distribution. The use of AIC is also illustrated in our example in Section~\ref{TreeApplication}.

\section{Simulation}\label{SimulationRegTree}
We have explored the performance of instability test for continuous partitioning variables and the performance of proposed SurvCART algorithm as a whole through simulation studies. In each of the following simulations both the survival times and censoring times were generated from exponential distribution under independent censoring. 
%
\subsection{Performance of instability test with continuous partitioning variable} \label {SimulationRegTree1}
This section is targeted to explore the size and power of parameter instability test for continuous variable as described in Section~\ref{InstabTest}. 

\subsubsection{Size of the test}\label{sizeoftest}
\begin{table}
\begin{center}
\caption{Size of proposed parameter instability test for continuous partitioning variable at 5\% level of significance via simulation as discussed in Section~\ref{sizeoftest}. The results are summarized based on $10,000$ simulations.}
\label{Sim1Size}
\begin{tabular}
{c|c|r|r|r|r|r|r}
\hline
  & & \multicolumn{6}{c}{Size (\%) of test}\\
\hline
$\lambda_T$ &Censoring rate& $N=50$ & $N=100$   & $N=200$ & $N=400$ & $N=1000$ & $N=2000$ \\
\hline
$1/20$ & 10\%&2.23&3.09&3.67&3.98&4.93&4.85\\
 & 25\%&2.76&3.09&3.97&4.23&4.73&4.65\\
 & 40\%&2.74&3.25&4.04&4.13&4.56&4.45\\
 & 60\%&3.14&3.46&3.55&4.58&4.47&4.60\\
\hline
$1/40$ & 10\%&2.19&3.06&3.66&3.94&4.94&4.83\\
          & 25\%&2.78&3.09&4.01&4.25&4.73&4.62\\
          & 40\%&2.67&3.31&4.04&4.18&4.51&4.47\\
          & 60\%&3.14&3.36&3.49&4.55&4.47&4.64\\
\hline
\end{tabular}
\end{center}
\end{table}

In order to examine the size of the test, survival times were generated from exponential distribution with hazard rate remain constant for entire population and independent of covariate value $X$. Precisely, survival times ($T^*$) and censoring times ($C$) for $N$ subjects were generated independently from exponential distributions with parameters $\lambda_T$ and $\lambda_C$, respectively. Follow-up time ($T$), were calculated as $\min{\{T^*, C\}}$. The size of test was explored for each of the combinations of  $\lambda_T$ ($1/20$ or $1/40$), censoring rate (10\% to 60\%) and $N$ ($50$ to $2000$). For each combination, $\lambda_C$ were determined from the following formula: $E(\delta)=\dfrac{\lambda_T}{\lambda_T + \lambda_C}$, where $E(\delta)$ indicates the expected censoring rate. In each simulation, the observations for covariate $X$ were generated from uniform(0, 10) for half of the patients and  from uniform(10, 20) for half of the remaining patients. For each combination,  $10,000$ replicates were generated, the test statistic $D(\lambda_T)$ (see Eq. \eqref{eq:D_T_exp}) were generated for each replicate and size of the test was determined as proportion of $D(\lambda_T)$ exceeds the $95$th percentile of its limiting distribution.\\

The size of the test for parameter instability test assuming exponential distribution for both $T^*$ and $C$ are summarized in Table~\ref{Sim1Size}. The size of the test approaches to the nominal significance level of 5\% with the increase in the sample size $N$, and it becomes very close to nominal level. The test is under-sized for smaller sample sizes; however, the reduced size has been also reported in other tests [e.g., Kolmogorov Smirnov test for normality] based on the Brownian Bridge process (\citealp{ massey1951kolmogorov, birnbaum1952numerical, lilliefors1967kolmogorov}). The censoring rate or event time distribution does not seem to influence the size of the test.

\subsubsection{Power}\label{InstabilityPower}
In this simulation, survival times were generated with hazard rate that varied with the covariate value $X$. We considered the population consists of following two subgroups with differential survival rate. Survival times ($T^*$) were generated from exponential distribution with hazard rate as $\lambda_{T1}$ and $\lambda_{T2}$, in subgroup 1 and subgroup 2, respectively. Censoring distribution were assumed same in both the subgroups; censoring times ($C$) were generated from $\mbox{exponential}(\lambda_C)$. Covariates values were generated from $\mbox{Uniform} (0, 10)$ in subgroup 1 and from $\mbox{Uniform} (10, 20)$ in subgroup 2. In each simulation replicate, $N1+N2$ observations were generated of which $N1$ values belongs to subpopulation 1 and remaining $N2$ values came from subpopulation 2. We set the value of  $\lambda_{T1}$ at $1/20$. The values of $\lambda_{T2}$ ($1/20$, $1/30$, $1/40$ or $1/60$), $\lambda_{C}$ ($1/30$, $1/40$, or $1/50$), $N1$ (25 to 400) and $N2$ (25 to 400) were varied in the simulation.  For each combination,  $10,000$ replicates were generated, the test statistic $D(\lambda_T)$ (see Eq. \eqref{eq:D_T_exp}) were generated for each replicate and power of the test was determined as proportion of $D(\lambda_T)$ exceeds the $95$th percentile of it limiting distribution .\\

\begin{table}
\begin{center}
\caption{Power (\%) of parameter instability test with continuous partitioning variable at 5\% level of significance via simulation described in Section \ref{SimulationRegTree1}. $N1$ and $N2$ represent number of subjects come from event time distribution with hazard rate $\lambda_1$ and $\lambda_2$, respectively.} \label{Sim1TreeSummary}
\begin{tabular}
{r|c|c|c|r|r|r|r|r|r|r}
\hline
&\multicolumn{2}{c|}{}&Censoring&\multicolumn{7}{c}{}\\
&\multicolumn{2}{c|}{Event rates}&rates&\multicolumn{7}{c}{Power (\%)}\\
\hline
&&&&N1=25&N1=25&N1=50&N1=50&N1=100&N1=200&N1=400\\
&$\lambda_{T1}$&$\lambda_{T2}$&$\lambda_{C}$&N2=25 &N2=25&N2=50&N2=75&N2=100&N2=200&N2=400\\
\hline
\#1&1/20&1/30&1/30&11.1&12.7&21.9&24.9&44.2&76.1&97.1\\
\#2&1/20&1/40&1/30&27.7&30.7&54.7&63.1&87.1&99.5&$>$99.9\\
\#3&1/20&1/50&1/40&49.4&56.1&84.9&90.7&99.3&$>$99.9&$>$99.9\\
\#4&1/20&1/60&1/50&69.7&76.8&96.3&98.7&$>$99.9&$>$99.9&$>$99.9\\
\hline
\end{tabular}
\end{center}
\end{table}
The observed power based on 10,000 simulation are displayed in Table \ref{Sim1TreeSummary}. Power of the test is improved as the difference between  $\lambda_{T1}$ and $\lambda_{T2}$ gets bigger and bigger. For example, even with sample size of 50, the observed power is close to 70\% when the median survival time is improved by 200\% (see scenario \#4 in Table~\ref{Sim1TreeSummary}). There is also gradual increase observed power of the test  with the increase in sample size.   However, the test is mildly conservative when sample size is small and $\lambda_{T2}$ is close to $\lambda_{T1}$.

\subsection{Performance of SurvCART algorithm for survival data}\label{SimulationRegTree2}
\begin{figure}
\centerline{\includegraphics [angle=0,width=160mm, height=60mm]{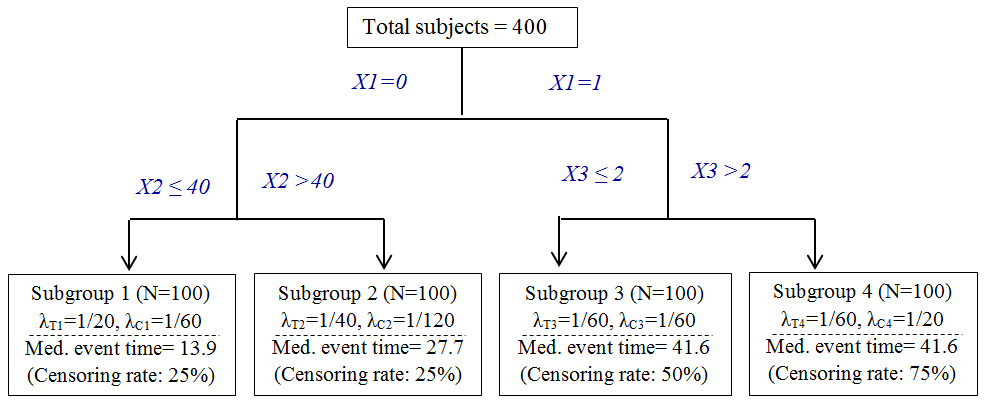}}
\caption{True tree structure for the simulation described in section~\ref{SimulationRegTree2}. There are only 3 subgroups when heterogeneity in censoring distribution is ignored - subgroup 1, subgroup 2 and combined subgroups of 3 and 4. When we have 4 subgroups when heterogeneity in both time to event and censoring distribution are considered.} \label{Sim2TrueTree}
\end{figure}

In this simulation, our goal is to assess the performance of SurvCART algorithm in comparison to other survival tree algorithms in truly heterogeneous population. Survival tree according to SurvCART algorithm were constructed assuming correctly and mis-specified time to event  distribution and with or without considering heterogeneity in censoring distribution. Performance of SurvCART algorithm was compared with the ctree algorithm (\citealp{hothorn2006unbiased}) for construction of survival tree, martingale residual based survival trees (\citealp{therneau1990martingale}) and relative-risk tree (\citealp{leblanc1992relative}). The survival tree according to ctree algorithm was obtained using \texttt{partykit} package in R (\citealp{hothorn2015partykit}). Martingale residual based survival tree and relative risk tree was obtained using \texttt{rpart} package in R (\citealp{therneau2010rpart}).  We did not consider other algorithm such as the RECPAM algorithm  (\citealp{ciampi1988recpam})  for comparison since no  R function  implementing these methods were available at the time of writing this article.\\

Data for each individual consisted of survival time, an indicator of censoring and three relevant covariates according to which event time and censor time distributions were assumed to vary: one dichotomous variable ($X1$) and two continuous variables ($X2\sim \mbox{uniform}(0, 100)$, $X3\sim \mbox{uniform}(0, 5)$). In addition, we also considered three nuisance covariates (i.e., related to neither event time  nor censoring time distributions): one continuous variable ($X4\sim \mbox{uniform}(0, 100)$),  one dichotomous variable ($X5$) with probability of 0.5 for success, and one categorical variable with 6 levels ($X6$) with equal probabilities for each categories. Overall the simulation was designed to generate data from a heterogeneous population with four subgroups characterized by covariates $X1$, $X2$ and $X3$ as displayed in Figure~\ref{SimulationRegTree2}. Of these 4 subgroups, subgroups 3 and 4 are similar in terms of time to event distribution, but are heterogeneous in terms of censoring distribution. That is, there are only 3 subgroups if we ignore heterogeneity in censoring distribution. However, when consider heterogeneity in both time to event and censoring distribution there are 4 subgroups in the true model.  In each subgroup, both the survival times and censoring times were generated from the respective exponential distribution with specfied $\lambda_T$ and $\lambda_C$ (Figure~\ref{SimulationRegTree2}).\\

We employed the following criteria for evaluating performance for each simulated dataset:
\[
\%\mbox{Difference from perfect tree}=\frac{\mbox{MAD}_{\mbox{fitted tree}}- \mbox{MAD}_{\mbox{perfect tree}}}{\mbox{MAD}_{\mbox{perfect tree}}}\times 100
\]
where $\mbox{MAD}_{\mbox{perfect tree}}$ and $\mbox{MAD}_{\mbox{fitted tree}}$ are the mean absolute deviation (MAD) estimated $\lambda$'s under ``perfect tree'' and fitted tree, respectively. We have introduced the notion of ``perfect tree'' to indicate a tree when all the subjects are classified perfectly according to the true tree structure (i.e. the tree structure displayed in Figure~\ref{Sim2TrueTree}). The mean absolute deviations (MAD) in $\lambda_T$ and $\lambda_C$ in $k$th subgroups of the dataset after fitting survival tree were calculated as follows:
\[
\mbox{MAD}(\hat{\lambda}_T)=\frac{1}{N_k}\frac{\sum_{j \in S_k}{|\lambda_{T,k}-\hat{\lambda}_{T,j}|}}{\lambda_{T,k}}
\qquad
\mbox{MAD}(\hat{\lambda}_C)=\frac{1}{N_k}\frac{\sum_{j \in S_k}{|\lambda_{C,k}-\hat{\lambda}_{C,j}|}}{\lambda_{C,k}}
\]
where $\lambda_{T, k}$ and $\hat{\lambda}_{T,j}$ are the estimated and true values of $\lambda$  in the $k$th subgroup.  $\hat{\lambda}_{T,j}$'s and $\hat{\lambda}_{C,j}$'s are the maximum likelihood estimate of $\lambda_T$ and $\lambda_C$, respectively, obtained from the subgroup it belonged to after fitting the survival tree. $S_k$ is the set of indices for all individuals  in the $k$th subgroup.\\

\begin{figure}[!p]
\centerline{%
\includegraphics [angle=0,width=110mm, height=190mm]{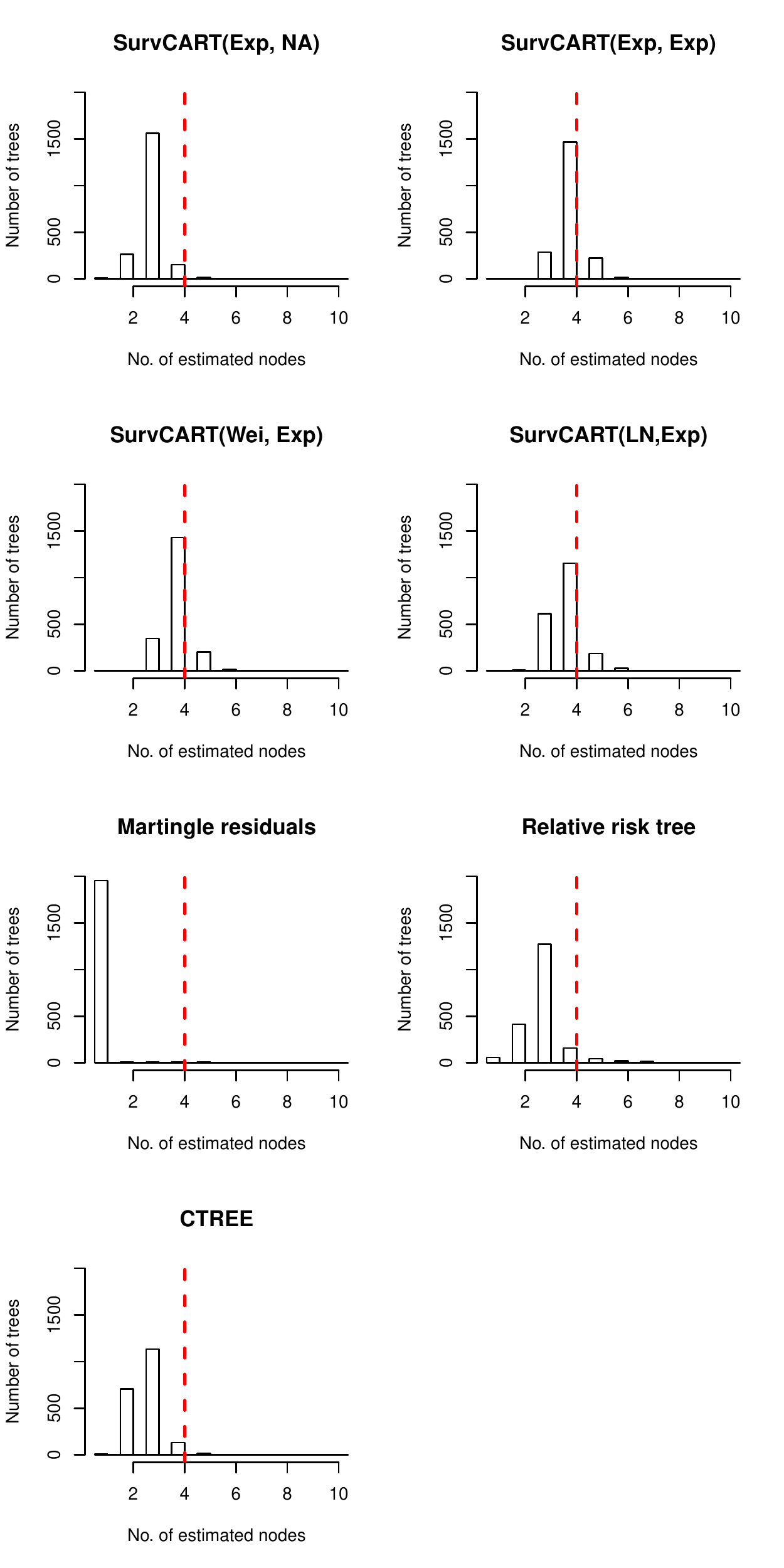}}
\caption{Number of tree nodes estimated by SurvCART algorithm and other tree fitting algorithms as described in Section \ref{SimulationRegTree2}. The dotted line indicates the true number of nodes equal to 4. SurvCART algorithms were fitted with specified distributions for $T^*$ and $C$. However, in SurvCART(Exp, NA) heterogeneity only in  distribution for  $T^*$ (but not in distribution of $C$) were considered.}
\label{Fig:Sim2hist}
\end{figure}

\begin{table}
\begin{center}
\caption{Comparison of SurvCART algorithm with the other tree fitting algorithms as described in Section \ref{SimulationRegTree2}} \label{Tab:Sim2ret}
\begin{small}
\begin{tabular}
{l|c|rrrrrrr|rr|rr}
\hline
&&\multicolumn{7}{c|}{Proportion(\%) of survival trees with } &\multicolumn{2}{c|}{Median }&\multicolumn{2}{c}{Median }\\
&Median &\multicolumn{7}{c|}{extracted subgroups} &\multicolumn{2}{c|}{ MAD($\lambda$)}&\multicolumn{2}{c}{ $\Delta$MAD($\lambda$)}\\
\cmidrule{3-9} \cmidrule{10-11} \cmidrule{12-13}
&nodes&1&2&3&4$^*$&5&$6-7$&$\ge 8$&$\lambda_T$&$\lambda_C$&$\lambda_T$&$\lambda_C$\\
\hline
Perfect tree$^1$&&&&&&&&&1.08&0.97&&\\
SurvCART(Exp, NA)&3&$<1$&13&78&7&$<1$&$<1$&&1.33&3.74&15.78&286.71\\
SurvCART(Exp, Exp)&4&&&14&73&11&$<1$&&1.49&1.16&25.30&9.48\\
SurvCART(Wei, Exp)&4&&$<1$&17&71&10&$<1$&&1.53&1.17&28.05&10.56\\
SurvCART(LN, Exp)&4&&$<1$&31&58&9&2&&1.87&1.23&53.74&19.76\\
ctree$^2$&3&$<1$&35&57&7&$<1$&$<1$&&2.05&3.97&64.68&294.11\\
Martingle  residuals$^3$&1&98&$<1$&$<1$&$<1$&$<1$&$<1$&$<1$&4.08&4.50&285.22&363.01\\
Relative Risk tree$^4$&3&3&21&64&8&2&2&$<1$&1.57&3.84&35.26&294.05\\

 \hline
 \multicolumn{13}{l}{Simulation results are based on 2,000 simulated datasets}\\
 \multicolumn{13}{l}{*True number of node was 4.}\\ 
 \multicolumn{13}{l}{MAD($\lambda$): absolute deviation in estimation of $\lambda$; $\Delta$MAD($\lambda$): \%Increase in MAD($\lambda$) from perfect tree.}\\
\multicolumn{13}{l}{$^1$When all the subjects are classified perfectly according to the the true tree structure.}\\
\multicolumn{13}{l}{SurvCART algorithms were fitted with specified distributions for $T^*$ and $C$. However, in SurvCART(Exp, NA) }\\
\multicolumn{13}{l}{heterogeneity only in  distribution for  $T^*$ (but not in distribution of $C$) were considered.}\\
\multicolumn{13}{l}{$^2$fitted with ctree function in \texttt{partykit} package.}\\
\multicolumn{13}{l}{$^3$fitted with \texttt{rpart} package; R-code: \texttt{rpart(martingle-residual $\sim$X1+X2+X3+X4+X5+X6)}}\\
\multicolumn{13}{l}{$^4$fitted with \texttt{rpart} package; R-code: \texttt{rpart(Surv(timevar, censorvar) $\sim$ X1+X2+X3+X4+X5+X6)}}
\end{tabular}
\end{small}
\end{center}
\end{table}

The simulation results comparing SurvCART with the other existing algorithms are summarized in Table  \ref{Tab:Sim2ret}, Table \ref{Tab:Sim2ret2} and Figure \ref{Fig:Sim2hist} based on 2,000 simulations. The SurvCART(Exp, NA) that considered only heterogeneity in time to event distribution (assuming exponential distribution), but not in censoring distribution, identified 3 subgroups in 78\%  case (Table  \ref{Tab:Sim2ret}, Figure \ref{Fig:Sim2hist}). This is consistent with the fact that there were only 3 subgroups when heterogeneity in censoring is ignored - subgroup 1, subgroup 2 and combined subgroup 3 and 4 (see Figure~\ref{Sim2TrueTree}). SurvCART(Exp, NA) also correctly identified $X1$ as the first splitting variable in 94\% cases and $X2$ as the second splitting variable in 86\% cases, consistent with the true model ignoring heterogeneity (Table \ref{Tab:Sim2ret2}). \\

Survival tree considering heterogeneity in both time to event and censoring distributions according to SurvCART algorithm  were fitted assuming exponential (SurvCART(Exp, Exp)), weibull (SurvCART(Wei, Exp)) and log normal (SurvCART(LN, Exp)) time to event distribution. In all these three cases exponential censoring distribution were considered. Note that time to event and censoring distribution are correctly specified in SurvCART(Exp, Exp). SurvCART(Wei, Exp) is also consistent with the true data generating mechanism given that exponential is a special case of weibull distribution. However, time to event distribution was mis-specified in SurvCART(LN, Exp). SurvCART(Exp, Exp), SurvCART(Wei, Exp) and SurvCART(LN, Exp) extracted 4 subgroups in 73\%, 71\% and 58\% cases (Table  \ref{Tab:Sim2ret}, Figure \ref{Fig:Sim2hist}). When comparing to other survival tree methods, both the ctree and relative risk tree algorithms extracts 3 subgroups on average. Unlike SurvCART algorithm, these two methods are designed to extract subgroups based on heterogeneity of only time-to-event distribution and probably due to this reason cannot distinguish the Subgroups 3 and Subgroups 4. Therefore, it would be more appropriate to compare ctree and relative risk tree method with SurvCART(Exp, NA). These two methods extracts exactly 3 subgroups in 57\% and 64\% cases respectively based on heterogeneity of time-to-event distribution in comparison to 78\% cases in SurvCART(Exp, NA). Interestingly, use of permutation based test in ctree algorithm does not make  any improvement when compared to ctree algorithm, possibly due to the well known conservatism of permutation based tests (\citealp{berger2000pros}).\\

In terms of metric MAD($\lambda$) and $\Delta$MAD($\lambda$), all four SurvCART algorithms seem to work better than the other tree methods indicating that subgroup identification is relative more accurate with SurvCART algorithm (Table  \ref{Tab:Sim2ret}). In terms of selection of splitting variable in SurvCART(Exp, Exp), SurvCART(Wei, Exp) and SurvCART(LN, Exp): $X1$ was selected as first splitting variable in 97\% of cases; $X2$ was selected as second splitting variable in 83\%, 80\% and 64\% cases, respectively; $X3$ was selected as second splitting variable in $>$98\% of cases. Most importantly, these three algorithms identified the correct tree (only accounting for splitting variable, but not cut-off point) in 81\%, 78\% and 63\% cases (Table  \ref{Tab:Sim2ret2}). All these results suggests that SurvCART algorithm performs relatively better than the other methods, even with mis-specified time to event distribution. However,  correct specification of underlying distribution indeed improves the performance of SurvCART algorithm.  In our simulation, we observed  the AIC of constructed tree (obtained from Cox model with subgroups as strata) is higher when underlying model is correctly specified (Table  \ref{Tab:Sim2ret2}). This suggests AIC can be appropriately used in choosing the underlying distribution to construct survival tree using SurvCART algorithm.  

\begin{table}
\begin{center}
\caption{Evaluation of performance SurvCART algorithm under mis-specification of time to event and censoring distribution as described in Section \ref{SimulationRegTree2}} \label{Tab:Sim2ret2}
\begin{small}
\begin{tabular}
{l|c|cccc}
\hline
& &\multicolumn{4}{|c}{Proportion (\%) of survival trees with}\\\cmidrule{3-6}
&Median&$X1$ as first&$X2$ as first or &$X3$ as first or&Selection of splitting\\
&AIC&  splitting&second splitting&second splitting&  variable identical\\
&(tree)$^1$& variable &  variable &variable &  to true tree\\
\hline
SurvCART(Exp, NA)&-1705.34&94.40 &85.65& 1.15 & 0.65 \\
SurvCART(Exp, Exp)&-1620.94& 97.20  & 82.55 &98.65 &81.10\\
SurvCART(Wei, Exp)&-1626.88& 96.95& 79.75 & 98.10&77.85\\
SurvCART(LN, Exp)&-1655.44& 97.05 & 63.90 & 98.45 & 62.75\\
 \hline
 \multicolumn{6}{l}{Simulation results are based on 2,000 simulated data sets}\\
\multicolumn{6}{l}{$^1$AIC from Cox regression model for tree structure (i.e., stratified Cox model with subgroups as strata).}\\
\multicolumn{6}{l}{SurvCART algorithms were fitted with specified distributions for $T^*$ and $C$. However, in SurvCART(Exp, NA) }\\
\multicolumn{6}{l}{heterogeneity only in  distribution for  $T^*$ (but not in distribution of $C$) were considered.}
\end{tabular}
\end{small}
\end{center}
\end{table}
%
\section{Application}\label{TreeApplication}
	We have applied SurvCART algorithm on the {\it recurrence free survival}  (RFS) time originated from prospective randomized clinical trial conducted by German Breast Cancer Study Group (GBSG)  (\citealp{schumacher1994randomized}). The purpose was  to evaluate the effect of prognostic factors on RFS time  among node positive breast cancer patients receiving chemotherapy in adjuvant setting. RFS was defined as time from mastectomy to the first occurrence of either recurrence, contralateral or secondary tumor, or death. The dataset was accessed from ipred package in R. The available dataset contains observation on RFS follow up time (with censoring status) from 686 breast cancer patients along with information on several prognostic variables including hormonal therapy (yes/no), age, menopausal status (Pre/post), tumor size, tumor grade (I/II/III), number of positive nodes, level of progesterone receptor (PR) and level of estrogen receptor. For details about the conduct of the study, please refer to Schumacher {\it et al.} (\citealp{schumacher1994randomized}).    The median RFS time based on entire 686 patients was 60 months with total of 299 reported RFS events. Subgroup analysis (\citealp{schumacher1994randomized})  and survival tree analysis using ctree algorithm (\citealp{hothorn2006unbiased}) were carried out on the data from this study earlier. \\
	
For the construction of survival tree we used all the above mentioned  prognostic variables as partitioning variables. We have fitted SurvCART algorithm assuming weibull, exponential and log-normal distributions for $T^*$ and $C$. SurvCART algorithm  was applied with the following specifications: (1) the significance level for individual instability test was set to 10\%, (2) the minimum node size for further split was set to 50, and (3) the minimum terminal node size was set to 25.\\

\begin{figure}
\includegraphics[angle=0,width=170mm, height=85mm]{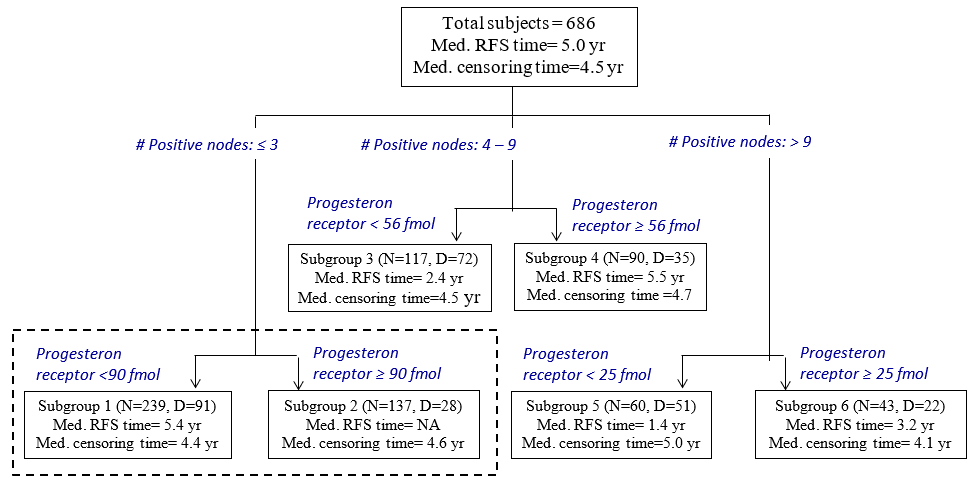}
\includegraphics[angle=0,width=170mm, height=85mm]{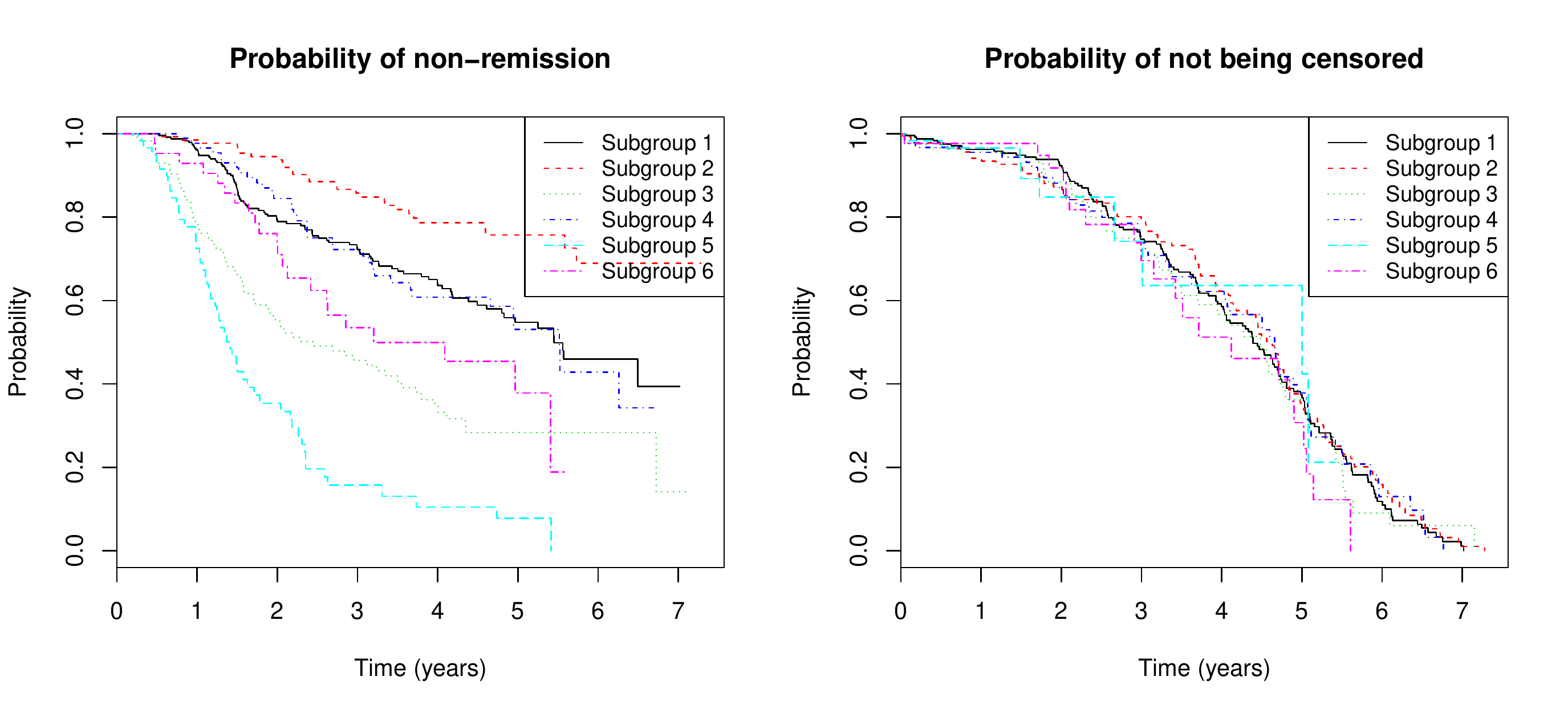}
\caption{{\it Top panel.} Estimated Survival tree obtained via SurvCART algorithm for RFS time as discussed in Section~\ref{TreeApplication}.  The split within the dotted line was observed with Weibull (or, log normal) time-to-event distribution but not with exponential time to event distribution. Exponential censoring distribution was considered. $N$ represents the number of patients and $D$ indicates number of events. Median RFS time and censoring times represent Kaplan-Meier estimates. {\it Bottom panel.} RFS probabilities (on {\it left})  and censoring probabilities (on {\it right}) for each subgroups separately.} \label{fig:RFStree}
\end{figure}

The survival tree was fitted assuming exponential, weibull and log normal distribution; however, the censoring distribution was kept fixed at exponential distribution. The SurvCART algorithm with exponential time to event distribution identified 5 subgroups. SurvCART algorithm with weibull or log normal as time to event distribution identified 6 subgroups - these subgroups are very consistent with the subgroups identified  with the exponential time-to-event distribution with only exception that one of the subgroup was further split into two groups (see Figure \ref{fig:RFStree}, top panel). AIC from Cox regression model for root node (i.e., Cox model without any covariate) was -3576.2.  AIC from Cox regression model for tree structure (i.e., stratified Cox model with subgroups as strata) were -2629.2 for 5 subgroups  and -2488.0 for 6 subgroups. Therefore, we present here the survival tree with 6 subgroups. This survival tree suggests that number of positive nodes and  PR level are key determinants for RFS time. The patients with reduced number of positive nodes at baseline experienced improved RFS time and the RFS time deteriorates with increase in number of positive nodes. Among the patients with 4 or more positive nodes, those with higher PR level experienced better RFS time. Based on these two prognostic factors, the tree identified 6 subgroups. These 6 subgroups are quite different in terms of RFS probabilities (see, Figure \ref{fig:RFStree}, bottom left panel). Patients with 3 or fewer positive nodes and PR level 90 fmol or more (i.e., subgroup 2) constitutes the best subgroup in terms of RFS time whereas patients with 9 or more positive nodes and PR level less than 25 fmol (i.e., subgroup 5)  are at higher risk of experiencing remission. In terms of censoring probabilities all the 6 subgroups looks very similar (see, Figure \ref{fig:RFStree}, bottom right panel). \\

Our findings are in consistent with the findings of schumacher et al. \cite{schumacher1994randomized}. They found number of positive nodes and PR level are the only significant prognostic factors that influence RFS.  In terms of number of positive nodes, they categorized patients into 3 subgroups ($\le 3$, $4 - 9$ and $> 9$)  which is exactly similar to what we have obtained via SurvCART algorithm. For PR level, they chose 20 fmol as cut-off value for the prognosis of RFS whereas the our survival tree suggest a varying cut-off for PR depending on number of positive nodes.  Findings of ctree algorithm are slightly different - it found hormonal therapy as important prognostic variables as well along with number of positive nodes and PR level (\citealp{hothorn2006unbiased}). Overall ctree algorithm identified 4 subgroups - (1) number of positive nodes $\le 3$ and without hormonal therapy, (2) number of positive nodes $\le 3$ and with hormonal therapy, (3) number of positive nodes $> 3$ and PR level $\le 20$ fmol and (4) number of positive nodes $> 3$ and PR level $> 20$ fmol.


\section{Discussion}\label{sec:disc}
The survival time distribution (and, also  censor time distribution) in a population may be influenced by one or more baseline covariates. In the context of medical research, this amounts to patient's survival may be influenced by prognostic variables. Survival tree offers an efficient tool to explore the influence of covariates on survival time and censor time distributions including the interaction effects of covariates. In fact, we have seen a plethora of recent research to identify subgroups using survival tree in many disease setting including coronary heart disease (\citealp{ramezankhani2017new}), kidney disease (\citealp{ramezankhani2017application}) and infectious disease (\citealp{yoon2017tree}), to name a few. We have proposed SurvCART algorithm to construct survival tree. Our proposed SurvCART algorithm has two major advantages: firstly, it identifies subgroups on the basis of both event time and censor time distribution and secondly, it selects splitting variable via formal statistical test. Because the splitting variables are selected via formal statistical test, it offers guard against selection bias and spurious finding.\\ 

The framework presented for SurvCART falls under conditional inference framework. In summary, the proposed SurvCART algorithms works as follows: (1)  in step 1, carry out the parameter instability test for each covariate to identify the best splitting variable, and (2) in step 2, choose the split point based on appropriately chosen maximal statistic. The parameter instability in step 1 only requires expression for score function and its variance. For the step 2, a dissimilarity measure is required and split point with maximum value of dissimilarity measure is used for growing the tree. Since the proposed survival tree algorithm is likelihood or score based, this framework can be extended in constructing tree in other set-up as well as long  expression for score function (with its variances) and dissimilarity measure available for steps 1 and 2, respectively. Thus in future effort will be made to extend this approach to construct survival tree for left truncated right censored data. Another interesting area in clinical trial (or medical research in general) to identify subgroups with distinct hazards ratio (HR), an indicator of treatment benefit. There is no known approach to-date to construct tree to identify subgroups with varying HR. The present framework can be extended towards that to construct tree based on varying HR.

\section{Software}
\label{sec5}

Software in the form of \verb|R| package \verb|LongCART| (\citealp{kundu2020longcart}) together with a sample input data set and complete documentation is available on CRAN. The parameter instability tests for categorical partitioning variables described in Section~\ref{sec:StabCat} and continuous partitioning variable described in Section~\ref{sec:StabCont} are implemented in \verb|StabCat.surv()| and \verb|StabCont.surv()|, respectively. The SurvCART algorithm to construct survival tree described in Section~\ref{Algorithm} is implemented in \verb|SurvCART()| through \verb|R| package \verb|LongCART| (\citealp{kundu2020longcart}) (see Appendix D for illustration).

\section*{Acknowledgments}

Author would like to acknowledge Prof. David Hosmer and Prof. Stanely Lemeshow for their generous permission to use German Breast Cancer Study data in the manuscript.

\section*{Appendices}
\subsubsection*{Appendix A: Test statistic for parameter instability test presented in Section~\ref{sec:StabCat}}
 The score functions $\mathbf{u}(\mathbf{y}_i, \hat{\boldsymbol{\theta}})$ are independent. Further, under $H_0$, $E_{H_0}[\mathbf{u}(\mathbf{y}_i, \boldsymbol{\theta}_0)]=0$ and $\mathbf{u}(\mathbf{y}_i, \hat{\boldsymbol{\theta}}) |_{H_0}\rightarrow^d N[0, \hat{\mathbf{J}}] $ where $\hat{\mathbf{J}}=\mathbf{J}(\hat{\boldsymbol{\theta}})$. Therefore, for a categorical partitioning variable, $X$, 
\begin{equation}\label{eq:LongCART1}
\chi^2_{cat}=\sum_{g=1}^G{\left[\sum_{i=1}^N{I(X_i=c_{(g)})\mathbf{u}(\mathbf{y}_i, \hat{\boldsymbol{\theta}})}\right]^{\top}\left[m_g\hat{\mathbf{J}}\right]^{-1}\left[\sum_{i=1}^N{I(X_i=c_{(g)})\mathbf{u}(\mathbf{y}_i, \hat{\boldsymbol{\theta}})}\right]}
\end{equation}
is asymptotically distributed as $\chi^2$ with $(G-1)\times\dim{(\boldsymbol{\theta})}$ degrees of freedom under $H_0$.  The reduction in $\dim{(\boldsymbol{\theta})}$ degrees of freedom is due to the estimation of $\boldsymbol{\theta}$ from the data. Now using C1--C3 (Section~\ref{PrelimLongTree}), we can decompose $\chi^2_{cat}$ in Eq.~\ref{eq:LongCART1} into independent components $\chi^2_T$ and $\chi^2_C$ specified in Eq.~\ref{eq:stabcatT} and Eq.~\ref{eq:stabcatC}, respectively.

\subsection*{Appendix B: Test statistic for parameter instability test presented in Section~\ref{sec:StabCont}}
Let's define following standardized estimated score process
\begin{equation}
 \mathbf{M}_N(t, \boldsymbol{\theta})=N^{-1/2}\mathbf{J}^{-1/2}(\hat{\boldsymbol{\theta}}) \sum_{i=1}^{M_g}{\mathbf{u}(\mathbf{y}_i, \boldsymbol{\theta})}
\end{equation}
As shown by \citet{kundu2019regression}, under $H_0$, each component of above process is asymptotically distributed as independent standard Brownian Bridge  processes. Now using C1--C3 (Section~\ref{PrelimLongTree}), 
\[
 \mathbf{M}_N(t, \hat{\boldsymbol{\theta}})=
 \left[
\begin{matrix}
 \mathbf{M}_N(t, \hat{\boldsymbol{\theta}}_T)\\
 \mathbf{M}_N(t, \hat{\boldsymbol{\theta}}_C)\\
\end{matrix}
\right]
\]
and hence, each component of $ \mathbf{M}_N(t, \hat{\boldsymbol{\theta}}_T)$ and $ \mathbf{M}_N(t, \hat{\boldsymbol{\theta}}_C)$ process is also asymptotically distributed as independent standard Brownian Bridge  processes under $H_0$. That is,
\begin{align}
M_N(t, \hat{\theta}_{T,q}) \rightarrow_d W^0(t)   \;\;\; q^{th}\;\; (q=1, \cdots, \dim{(\boldsymbol{\theta}_T)}) \\
M_N(t, \hat{\theta}_{C,r}) \rightarrow_d W^0(t)   \;\;\; r^{th}\;\; (q=1, \cdots, \dim{(\boldsymbol{\theta}_C)})
\end{align}

The above weak convergence continues to hold for any `reasonable' functional (including supremum) of  $M_N(t, \hat{\theta}_k)$. Therefore, the quantities $D(\theta_{T,q})$ in  Eq.~\ref{eq:D_T} and $D(\theta_{T,q})$ in Eq.~\ref{eq:D_C} are independently distributed as supremum of standard Brownian Bridge process with known distribution function as specified in \eqref{eq:D_dist}(\citealp{billingsley2009convergence}).

\subsection*{Appendix C: Score function under log-normal and normal distributions}
First, consider the log-normal distribution with $f(t;\mathbf{\boldsymbol{\theta}}_{T})=\mbox{LN}(\mu_T, \sigma_T)$ and $g(c; \mathbf{\boldsymbol{\theta}}_{C})=\mbox{LN}(\mu_C, \sigma_C)$

Let's, define:
\[
y_i=\frac{\log{t_i}-\mu_T}{\sigma_T} \qquad 
h(y)=\frac{\phi(y)}{\Phi(-y)}
\]
where, $\phi(y)$ and $\Phi(y)$ are the density function and cumulative distribution function from standard normal distribution. Then, we have, the score function as follows (e.g., see \citealp{kundu2007hybrid}): 
\begin{equation*}
\mathbf{u}_i(\boldsymbol{\theta}_T)=
\left[
\begin{matrix}
u_i(\mu_T)\\
u_i(\sigma_T)\\
\end{matrix}
\right]
=\frac{1}{\sigma_T}
\left[
\begin{matrix}
\delta_i y_i + (1-\delta_i )h(y_i)\\
\delta_i (y_i^2-1)+ (1-\delta_i )y_i h(y_i)\\
\end{matrix}
\right]
\end{equation*}

Similarly, $\mathbf{u}_i(\boldsymbol{\theta}_C)$ and $\mathbf{J}(\boldsymbol{\theta}_C)$ can be obtained redefining $y_i=(\log{t_i}-\mu_C)/\sigma_C$ and replacing $\mu_T$, $\sigma_T$ and $\delta_i$ by $\mu_C$, $\sigma_C$ and $(1-\delta_i)$, respectively.\\

For normal distributions with $f(t;\mathbf{\boldsymbol{\theta}}_{T})=\mbox{N}(\mu_T, \sigma_T)$ and $g(c; \mathbf{\boldsymbol{\theta}}_{C})=\mbox{N}(\mu_C, \sigma_C)$, the expression for score function is almost same with only exception that $\log(t_i)$ should be replaced by $t_i$.

\subsection*{Appendix D: R code to construct survival tree}

\begin{small}
\verb|library(LongCART)|

\verb|data(GBSG2)|\\

\verb|#--- numeric coding of character variables|

\verb|GBSG2$horTh1= as.numeric(GBSG2$horTh)|

\verb|GBSG2$tgrade1= as.numeric(GBSG2$tgrade)|

\verb|GBSG2$menostat1= as.numeric(GBSG2$menostat)|\\

\verb|#--- Add subject id|

\verb|GBSG2$subjid= 1:nrow(GBSG2)|\\

\verb|#--- Run SurvCART() with time-to-event distribution: weibull, censoring distribution: exponential|
  
\verb|out= SurvCART(data=GBSG2, patid="subjid", censorvar="cens", timevar="time",| 

\hspace{0.3in}        \verb|gvars=c('horTh1', 'age', 'menostat1', 'tsize', 'tgrade1', 'pnodes', 'progrec', 'estrec'),  |
        
\hspace{0.3in}         \verb|tgvars=c(0,1,0,1,0,1, 1,1), time.dist="weibull", cens.dist="exponential", |  
              
\hspace{0.3in}         \verb|event.ind=1,  alpha=0.05, minsplit=80, minbucket=40, print=TRUE)|\\

\verb|#--- Plot tree|

\verb|par(xpd = TRUE)|

\verb|plot(out, compress = TRUE)|

\verb|text(out, use.n = TRUE)|\\

\verb|#--- Plot KM plot of event times for subgroups identified by tree|

\verb|KMPlot.SurvCART(out, scale.time=365.25, type=1)|
\end{small}

\end{document}